\documentclass[%
 reprint,
 superscriptaddress,
 amsmath,amssymb,
 aps,
pre,
floatfix,
a4paper]{revtex4-2}

\usepackage{amsmath,amssymb}
\usepackage{fix-cm}
\DeclareMathAlphabet{\mathsfit}{T1}{\sfdefault}{\mddefault}{\sldefault}
\SetMathAlphabet{\mathsfit}{bold}{T1}{\sfdefault}{\bfdefault}{\sldefault}

\usepackage{graphicx}
\usepackage{dcolumn}
\usepackage{xcolor}
\usepackage{bm}
\usepackage[colorlinks,citecolor=blue]{hyperref}
\usepackage[export]{adjustbox}
\usepackage[T1]{fontenc}

\newcommand{\dd}[0]{\mathrm{d}}
\newcommand{\pd}[0]{\partial}
\newcommand{\mm}[1]{\bm{\mathsf{#1}}}
\newcommand{\changed}[1]{{\leavevmode#1}}

\begin{document}

\title{The trimer paradox: the effect of stiff constraints on equilibrium distributions in overdamped dynamics}

\author{Radost Waszkiewicz}
\email{radost.waszkiewicz@gmail.com}
\affiliation{Institute of Physics, Polish Academy of Sciences, Aleja Lotników 32/46, PL-02668 Warsaw, Poland}

\author{Maciej Lisicki}
\email{mklis@fuw.edu.pl}
\affiliation{Institute of Theoretical Physics, Faculty of Physics, University of Warsaw, L. Pasteura 5, 02-093 Warsaw, Poland}

\date{\today} 

\begin{abstract}
    We reconsider the classical problem of a freely joined chain of Brownian particles connected by elastic springs and study its conformational probability distribution function in the overdamped regime in the limit of infinite stiffness of constraints.
    We show that the well-known solution by Fixman [Proc. Natl. Acad. Sci. USA {\bf 71}, 3050 (1974)] is missing a shape-related term, later alluded to but not computed by Helfand [J. Chem. Phys {\bf 71}, 5000 (1979)].
    In our approach, the shape term, also termed zero-point energy, arises explicitly from a careful treatment of the distributional limit.
    We present a computationally feasible method for the calculation of the shape term and demonstrate its validity in a couple of examples.
\end{abstract}

\maketitle

\section{Introduction}

Molecular Dynamics (MD) and Brownian Dynamics (BD) simulations are now standard tools for detailed modelling of a plethora of molecular and mesoscopic systems, where structural complexity poses a challenge to theoretical calculations. Upon the introduction of a suitable coarse-graining scheme to represent the composition of a given molecule, its conformational space is prescribed by specifying intramolecular interactions, such as chemical bonds or electrostatic forces, between the subunits. The nature of these interactions endows molecular models with elasticity, which in turn affects their conformational variability, as well as their macroscopic and statistical properties, such as rheology, diffusivity, and thermodynamics, and also biological function. 

A popular idealisation for a wide variety of systems is purely mechanical, where the molecule is represented by beads connected with springs, an idea which probably originated from Kramers~\cite{Kramers_1946}, who built on concepts of Kuhn, and termed it the pearl-necklace model. Variations of this idea are now the cornerstone of polymer physics~\cite{Flory_1953}, but have also been used to describe other classical many-body systems. Physically, the introduction of intramolecular interactions leads to constrained dynamics with multiple time scales. Typically, the characteristic time scales associated with internal vibrations of molecular bonds are much shorter than those of translational and rotational motion or bond angle dynamics. The disparity in the time scale of relaxation of hard (or stiff) degrees of freedom (such as bond lengths) compared to soft degrees of freedom (such as bond angles) is a source of stiffness in the problem and hinders fast numerical simulation of such systems~\cite{Frenkel_2023}. 

In many applications, the dynamics of the hard degrees of freedom are of secondary importance, and one method of circumventing this difficulty is to treat them as fully constrained (i.e. rigid), thus eliminating the short timescale, and enabling faster simulations. This is the basis of many algorithms, such as SHAKE~\cite{Ryckaert_1977} and its further extensions. During the development of such rigid models, a disconnect was discovered between the equilibrium bond angle distributions in the rigid simulations and the simulations with bonds modelled as very stiff springs. \changed{Further summary of literature on the topic is discussed in Sec.~\ref{sec:stiff_vs_rigid}.} 

The simplest example in which the complexity of the constrained dynamics can be appreciated is the case of a flexible trimer: a hypothetical molecule containing 3 subunits and 2 harmonic bonds, depicted in Fig. \ref{fig:trimer}a. The angle between the two bonds is denoted by $\psi$. In the classical textbook~\cite{Frenkel_2023}, Frenkel and Smit present an apparently paradoxical result: the marginal distribution of the bond angle in the case of rigid bonds, $\dd p_\text{rigid}$, and the limiting distribution of spring-like bonds where the spring stiffness $k$ is taken to infinity, $\dd p_{k\to\infty}$, do not coincide.
More precisely, they report
\begin{eqnarray}
    \dd p_{k\to\infty} & = & N_1 \sin \psi \,\, \dd \psi, \label{eqn:trimer-soft}                            \\
    \dd p_\text{rigid}      & = & N_2 \sin \psi \sqrt{{1-\frac{\cos^2\psi}{4}}} \,\, \dd \psi, \label{eqn:trimer-rigid}
\end{eqnarray}
for appropriate normalisation constants $N_1$ and $N_2$. Qualitatively, the distribution~\eqref{eqn:trimer-soft} corresponds to a spherically uniform distribution of the second bond direction when the coordinates are aligned with the first bond, while in Eq.~\eqref{eqn:trimer-rigid} it seems that an additional force resists bond alignment. In an early and widely cited work, Fixman~\cite{Fixman_1974} presented a derivation to argue that the ratio of the two probability densities, Eqs.~\eqref{eqn:trimer-soft} and \eqref{eqn:trimer-rigid}, can be computed with the knowledge of the constraining surfaces alone and does not require knowledge of the shape of the confining potential. This presupposes that both of these distributions are well defined whenever the constraint is given, which we argue is not the case.

In this contribution, we present a clear mathematical procedure leading to the \changed{$\dd p_{k\to \infty}$} limit in the overdamped regime when inertial effects can be neglected. We show that the distribution given by Eq.~\eqref{eqn:trimer-soft} is not completely determined by fixing the bond lengths. To correctly predict the limiting distribution, knowledge about the nature of confinement is needed, and its structure survives in the final expression. Furthermore, we demonstrate that Fixman's expressions for the stiff limit miss a term, which is important even in the case of harmonic springs.

\section{Stiff vs. rigid constraints}

\label{sec:stiff_vs_rigid}%
The first calculations of equilibrium properties of flexible polymers can be traced to the papers of Kramers~\cite{Kramers_1946}, Gō and Sheraga~\cite{Go_1969} and Fixman~\cite{Fixman_1974}. The results of the latter paper were used essentially without changes in later works, such as~\cite{Pear_1979, Hinch_1994, Otter_1998, Liu_1989}. Early papers focused on the treatment of constraints in the Lagrangian and Hamiltonian pictures, both classical and quantum-mechanical, by defining the soft and stiff degrees of freedom and integrating out generalised coordinates corresponding to the hard (stiff) variables. The limiting distribution is then obtained for infinitely stiff constraints and the stiff variables are eliminated.  

Fixman~\cite{Fixman_1974} presented a purely classical calculation of the limiting distribution, correctly writing the action of the distribution on a test function but erroneously assuming that the constraining potential depends only on the hard coordinates. As a result, the details of the confining potential cannot appear in his final expressions, which contain only the projected phase-space volume element terms. Gō and Sheraga~\cite{Go_1969} treated the problem quantum-mechanically and correctly identified two contributions to the equilibrium probability density of the soft coordinates: the projected volume elements and the zero-point energy of the vibrational motion of the hard coordinates. However, they then stated that the \changed{zero-point-energy} contribution in typical molecular systems varies only a little when changing the soft coordinates and may be altogether neglected. We show in the following that this is not necessarily the case in overdamped dynamics. By focusing on a concrete classical two-dimensional example with one constraint, Helfand~\cite{Helfand_1979} correctly identified the rigid-rod type distribution as a uniform distribution on the constraining manifold and computed the stiff-spring limit in a similar simple case. Unfortunately, these results are not expressed in a form applicable to higher-dimensional examples with multiple constraints. 

Due to this lack of generality, later works
cite Helfand as a reference to an erroneous claim that the bond angle distribution becomes non-uniform in the case of a rigid-rod system. The pitfall lies in the assumption that the distribution should be uniform under the action of a rotation matrix on a single bond, presumably by analogy with a dimer. However, there is a distinct difference between these two cases; In the case of a dimer, the true physical symmetries of the equations of motion -- rigid body rotation of the whole system and translations -- act transitively on the space of all possible configurations, and thus the equilibrium distribution can be determined from symmetry arguments. More generally, for a rigid body, the rotation group $\mathrm{SO}(3)$ can be treated as a topological manifold, resulting in a natural notion of uniformity, with the uniform distribution given by the Haar measure. 
For the trimer, however, we have configurations that are meaningfully distinct. 
There is no symmetry that changes the bond angle $\psi$. Although the equilibrium distribution for 6 out of 7 degrees of freedom can be determined from the six symmetries of translation and rotation, the relative weight of configurations with different $\psi$ has to refer to the physical problem, not just geometric considerations.

Hinch \textit{et al.}~\cite{Hinch_1994,Grassia_1995} approached the problem differently, by manipulating the Langevin equations, which are notoriously difficult to handle, and erroneously assumed that details of the confining potential are of little importance to the equilibrium distribution of the constrained configurations.
Van Kampen and Lodder~\cite{van_Kampen_1984} noted that the approach of Helfand~\cite{Helfand_1979} is generally applicable to constrained systems and comment that the discrepancy between the stiff-spring and rigid-rod distributions in the case of the trimer molecule is due to the fact that \emph{the width of this gully is not the same everywhere}~\cite{van_Kampen_1984}. However, they did not quantify the effect of this width in a higher-dimensional case. As a result, the knowledge of the influence of the shape of the constraining potential appears to have been lost as multiple works (for example,~\cite{Vitalis_2009, Allen_2004, Frenkel_2023}) mention only the densities of the volume elements and omit the \emph{zero point energy} or, equivalently, the potential \emph{shape} terms from their descriptions of constrained dynamics, widely using the result of Fixman~\cite{Fixman_1974} as a reference.

In the following, we show a way to compute the limiting distribution by integrating out the \changed{confined} degrees of freedom in the overdamped limit. We explicitly find the metric contribution arising from the transformation of coordinates and the shape term that represents the details of the constraining potential and which remains imprinted on the configurational distribution when the hard coordinates are taken to be infinitely stiff. 

\section{The stiff spring limit}

\begin{figure}[htb]
    \centering
    \includegraphics[width=\linewidth]{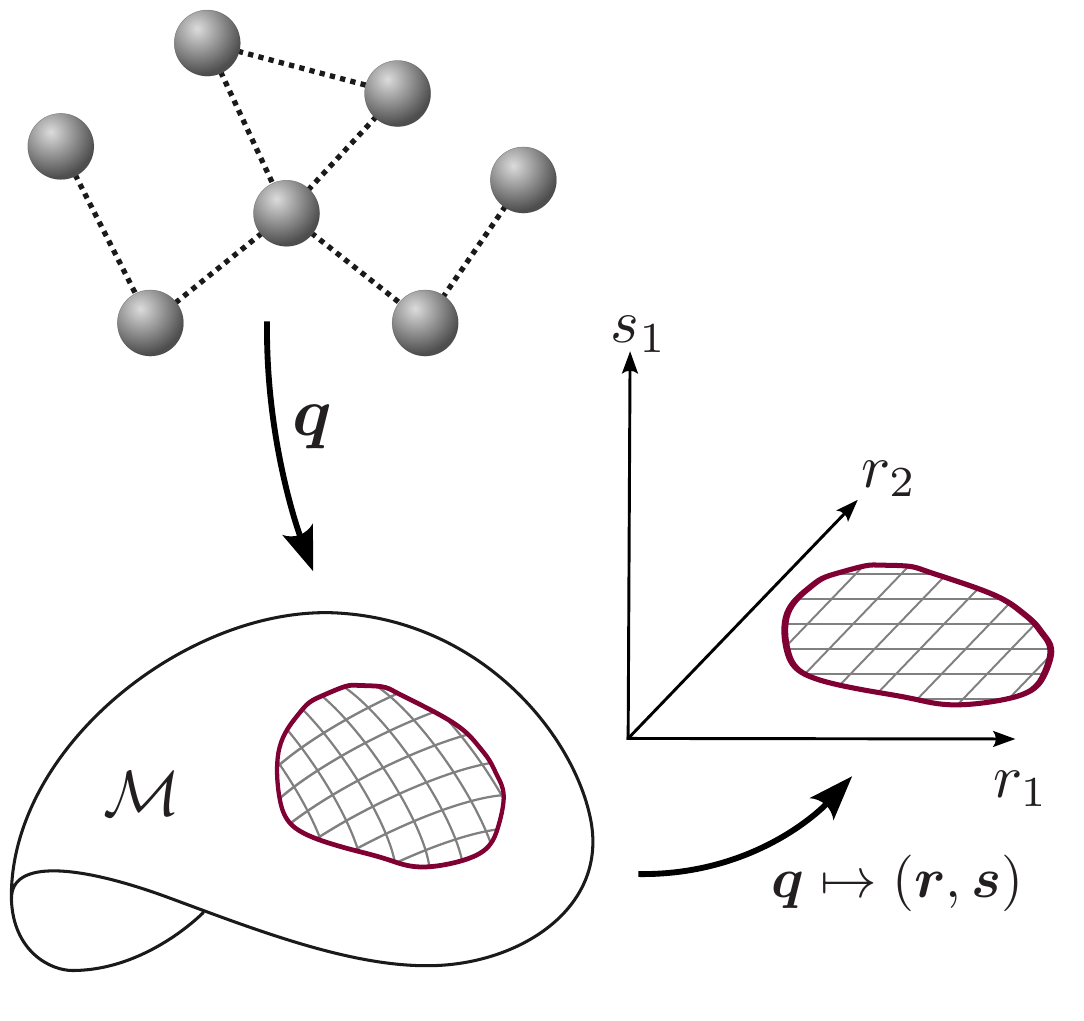}
    \caption{A flexible polymer with $N$ degrees of freedom is modelled as beads connected with springs.  Conformations of the polymer are described by a vector $\bm{q}$. The potential energy of the confining springs, $k^2 W(\bm{q})$, attains a minimum on a submanifold $\mathcal{M}$. 
    We define an orthonormal system of  coordinates on a compact neighbourhood of $\mathcal{M}$, enclosed by the red solid line, by introducing soft degrees of freedom $\bm{r}$ on $\mathcal{M}$ and hard degrees of freedom $\bm{s}$, which are normal to $\mathcal{M}$ and correspond to the confining springs. The hard degrees of freedom will be integrated out to find the limiting distribution for stiff springs.}
    \label{fig:sketch}
\end{figure}

\changed{This work focusses on the distributional limit of equilibrium distributions at increasingly stiff confinement: \emph{the stiff spring limit}.} 
We begin our consideration with a careful definition of \changed{this limit}. \changed{The rigid rod distribution is discussed only insofar as it is needed for the ``ratio of densities'' argument presented by \citet{Frenkel_2023}}.

We present a derivation of the limiting distribution in the following setting, sketched in Fig.~\ref{fig:sketch}. Consider a molecule (or a polymer) with a handful of identical subunits (atoms, beads, or monomers) each of mass $m$ and a total of $2N$ degrees of freedom ($N$ configurational and $N$ momentum) in a heat bath at constant temperature $T$, with $\beta = 1 / k_B T$. The conformation of the molecule is described by an $N$-dimensional vector $\bm{q}$ and the velocities of monomers are specified by a vector $\bm{v}$. We parametrise the conformations by Cartesian coordinates of monomers $q_i$ and Cartesian components of velocities $v_i$ with $i \in 1 \dots N$. The potential energy of the molecule has two components: the conformation-dependent energy $U(\bm{q})$ and the confining (springs) potential $k^2 W(\bm{q})$, where $k$ is a large parameter describing the spring stiffness. Note that $k$ here would correspond to the square root of a harmonic spring constant. The probability density $\hat{p}_k$ of the equilibrium distribution of the system configuration with respect to the Lebesgue measure $\dd \bm{q} \dd \bm{v}$ is given by the Maxwell-Boltzmann distribution \cite{Landau_2013}
\begin{equation}
    p_k = \hat{N}(k) \exp\left\{-\beta \left[U(\bm{q}) + k^2 W(\bm{q}) + \frac{1}{2} m \bm{v}\cdot\bm{v}\right]\right\},
    \label{eqn:maxwell-boltzman}
\end{equation}
with an appropriate normalising function $\hat{N}(k)$.
In the case under consideration $m$ does not depend on $\bm{q}$ so the momentum coordinates are easy to integrate out and we arrive at the familiar Boltzman probability density $p_k$ with respect to the measure $\dd \bm{q}$ \cite{Landau_2013} written as
\begin{equation}
    p_k = N(k) \exp\left\{-\beta \left[U(\bm{q}) + k^2 W(\bm{q}) \right]\right\},
    \label{eqn:boltzman}
\end{equation}
with a (different) normalising function $N(k)$. Crucially, this density is given with respect to canonical conformation coordinates and not just any coordinates -- if generalised coordinates are used, more care is needed and volume element density terms appear in expressions analogous to Eq.~\eqref{eqn:boltzman}. Moreover, in the case of generalised coordinates, the expression for kinetic energy is also more complicated. Clearly, Eq.~\eqref{eqn:boltzman} is a physical object and we are free to work in any coordinates of choice, and for the purpose of the derivation of the \emph{stiff spring limit} Cartesian coordinates are particularly convenient. We return to the treatment of generalized coordinates in \changed{later in this paper}.

With these caveats outlined, we define \emph{the stiff spring limit} $\dd p_{k\to\infty}$ as a measure which satisfies 
\begin{equation}
    \dd p_{k\to\infty} = \lim_{k\to\infty} \dd p_k,
\end{equation}
in the weak sense of the limit.

Suppose now that the spring potential is non-negative, $W(\bm{q}) \geq 0$, and attains a minimum for a configuration lying on a smooth submanifold $\mathcal{M}$ of dimension $M<N$, so that $W(\bm{q}) = 0 \iff \bm{q} \in \mathcal{M}$.
To determine the weak limit of $p_k$, we consider integrals $I_k$ for a compactly supported smooth test function $\phi(\bm{q})$ given by
\begin{equation}
    I_k = \int p_k(\bm{q}) \phi(\bm{q}) \dd\bm{q}.
\end{equation}
Since $\mathcal{M}$ is smooth, we can use the tubular embedding theorem to define new orthonormal coordinates $\bm{q} \mapsto (\bm{r},\bm{s})=(r_1,r_2,\dots,r_M,s_1,s_2,\dots)$ in the vicinity of $\mathcal{M}$, where $\bm{q} \in \mathcal{M} \iff \bm{s} = 0$. In other words, coordinates $r_i$ represent the \emph{soft} degrees of freedom and $s_i$ represent the \emph{hard} directions. Since the new coordinates are orthonormal, we know that the determinant of the transformation $(q_i) \mapsto (r_i,s_j)$ equals 1.
Additionally, $I_k$ is finite by the compactness of the support of $\phi$ and the continuity of $p_k$, so by Fubini's theorem we can replace the volume integral by an iterated one. Then, $I_k$ is given by
\begin{equation}
    I_k = \iint p_k(\bm{r},\bm{s}) \phi(\bm{r},\bm{s}) \dd\bm{s} \dd\bm{r}.
\end{equation}
We proceed by expanding $p_k$ in a Taylor series with respect to $\bm{s}=(s_1,s_2,\dots)$ to second order as $W(\bm{r},\bm{s}) = \bm{s}^{T}\mm{H}(\bm{r})\bm{s}$, where $\mm{H}(\bm{r})$ is the Hessian of $W$, evaluated at $\bm{s}=0$. Note that the linear term vanishes, since at $\bm{s}=0$ the potential $W$ attains a minimum. We can thus write
\begin{widetext}
\begin{equation}
        I_k = \iint   N(k) \exp\left\{ -\beta U(\bm{r},\bm{s}) -  \beta k^2 \left[ \bm{s}^T \mm{H}(\bm{r}) \bm{s} + o(||\bm{s}||^2) \right] \right\} \phi(\bm{r},\bm{s}) \dd \bm{s} \dd \bm{r}.
\end{equation}
     We now make a substitution $k\bm{s} = \bm{t}$ to arrive at
    \changed{\begin{equation}
        I_k = \iint \frac{N(k)}{k^{N-M}} \exp\left\{ -\beta U(\bm{r},k^{-1}\bm{t}) - \beta \left[ \bm{t}^T \mm{H}(\bm{r}) \bm{t} + ||\bm{t}||^2 \frac{ o(k^{-2} ||\bm{t}||^2)}{ k^{-2} ||\bm{t}||^2} \right] \right\} \phi(\bm{r},k^{-1}\bm{t}) \dd \bm{t} \dd \bm{r}.
    \end{equation}}
    Whenever $W$ increases sufficiently fast with the distance from the constraining manifold, we may take a limit inside the integral by the dominated convergence theorem to arrive at the limiting value $I = \lim_{k\to\infty} I_k$ given by
    \changed{\begin{equation}
        I = \iint \left(\lim_{k\to\infty} \frac{N(k)}{k^{N-M}} \right) \exp\left[ -\beta U(\bm{r},\bm{0}) - \beta\,  \bm{t}^T \mm{H}(\bm{r}) \bm{t} \right] \phi(\bm{r},\bm{0}) \dd \bm{t} \dd \bm{r},
    \end{equation}}
\end{widetext}
and, since $\mm{H}$ is full rank, perform the integral over $\bm{t}$ to arrive at
\begin{equation}
    I = L \int |\mm{H}(\bm{r})|^{-1/2} \exp\left[ -\beta U(\bm{r},\bm{0})         \right] \phi(\bm{r},\bm{0}) \dd \bm{r},
\end{equation}
with the constant
\changed{\begin{equation}
    L = \lim_{k\to\infty} \frac{N(k) \sqrt{(2\pi \beta)^{N-M}}}{k^{N-M}},
\end{equation}}
being independent of the soft coordinates.

In a particular physical setting, finding the orthonormal coordinates $(\bm{r},\bm{s})$ may be prohibitively difficult, but we can relax this strict requirement by considering another parametrisation $\bm{w}$ such that $\bm{q} = \bm{\zeta^*}(\bm{w})$, and that still separates \emph{soft} from \emph{hard} degrees of freedom by
\begin{equation}
    \bm{q} \in \mathcal{M} \iff \forall i > M : w_i = \xi_i,
\end{equation}
for a set of constants $\xi_i$. This parametrisation is not necessarily orthonormal. For convenience, we define a map $\bm{\zeta}: \mathbb{R}^M\to\mathbb{R}^N$ such that $\bm{\zeta}(w_1,w_2,\dots,w_M) = \bm{\zeta^*}(w_1,w_2,\dots,w_M,\xi_{M+1},\xi_{M+2},\dots,\xi_{N})$ which takes values of the \emph{soft} coordinates and returns points on $\mathcal{M}$. We now change variables under the integral from $\bm{r}$ to $\bm{w}$ to arrive at
\begin{equation}
        I = L\int   \frac{\left|\mm{J}^T \mm{J}\right|^{1/2}}{\left|\mm{H}(\bm{\zeta})\right|^{1/2}}  \exp\left[ -\beta U(\bm{\zeta})         \right] \phi(\bm{\zeta})  \dd w_1 \dots \dd w_M,
\end{equation}
where $\mm{J}$ is the Jacobian of $\bm{\zeta}(\bm{w})$. Note that $\bm{\zeta} : \mathbb{R}^M \to \mathbb{R}^{N}$, so the Jacobian is not a square matrix. We can express the above using the Dirac-$\delta$ distributions.
Taking
\begin{equation}
    \delta(\bm{w}' - \bm{\xi}') = \prod_{i = M+1}^{N} \delta(w_i - \xi_i),
\end{equation}
the limiting integral is recast as
\begin{equation}    
        I = L\int  \frac{\left|\mm{J}^T \mm{J}\right|^{1/2}}{|\mm{H}|^{1/2}} \exp \left( -\beta U \right)  \phi(\bm{\zeta^*}(\bm{w})) \delta(\bm{w}' - \bm{\xi}') \, \dd \bm{w}    
\end{equation}
for an appropriate constant $L$, and since $\phi$ is arbitrary, we arrive at $p_\infty = \lim_{k\to\infty} p_k$ such that
\begin{equation}
        \dd p_\infty = L  \frac{|\mm{J}^{T} \mm{J}|^{1/2}}{\left|\mm{H}\right|^{1/2}}  \exp \left[ - \beta U(\bm{\zeta^*}(\bm{w})) \right] \delta(\bm{w}' - \bm{\xi}') \ \dd \bm{w},
\end{equation}
in the weak sense.
We have thus arrived at the central result of this \changed{paper}.

The limiting distribution comprises two important factors. First, the metric term $|\mm{J}^{T} \mm{J}|$ describing the projection of the surface element of $\mathcal{M}$ onto $\mathbb{R}^{M}$ in the parametrisation $\bm{\zeta}(\bm{w})$; This term was correctly computed by Fixman~\cite{Fixman_1974}. Second, the \emph{shape} term $|\mm{H}|$, also called the zero-point energy in the quantum-mechanical setting, which was often missing in the derivations.

\changed{\section{Calculation of the shape term}}

A careful choice of parametrisation allows one to calculate the metric term with relative ease. However, the calculation of the Hessian term $|\mm{H}|$ may be more difficult. In this Section, we present a feasible approach to its determination.

Let $\mm{B}$ be the full Hessian of $W$ at a point $\bm{q}$,
\begin{equation}
    \mm{B}_{ij} = \frac{\pd^2}{\pd q_i \pd q_j} W(\bm{q}).
\end{equation}
Since the eigenvectors of $\mm{B}$ are orthogonal and zero eigenvectors lie inside the tangent space $T \mathcal{M}$, we can calculate $|\mm{H}|$ from the product of the non-zero eigenvalues of $\mm{B}$.
We can use knowledge of the tangent space of $\mathcal{M}$ to find the eigenvalues of $\mm{B}$.
If the confining function is of the form $W(\bm{q}) = \sum_i P_i^2$, we can find vectors that lie in the normal space $\mathcal{M}^{\perp}$~\cite[proposition 8.15]{Lee_2003} by calculating
\begin{equation}
    \mm{A}_{ji} =  \frac{\pd}{\pd q_j}
    P_i,
    \label{eq:matrix_a_defn}
\end{equation}
The vectors $\{\mm{A}_{j1},\ldots,\mm{A}_{j,N-M}\}$  are not pairwise orthogonal, but they are orthogonal to $\mathcal{M}$, and thus can form a basis of the normal space at $\bm{q}$.
We can write the eigenvalue problem using an arbitrary vector $\bm{b}$ as
\begin{equation}
    \lambda  \mm{A}_{ki} b_i = \mm{B}_{kj} \mm{A}_{ji} b_i.
\end{equation}
This is a system of $N$ equations with $N-M$ unknowns.
We can eliminate redundant equations by contracting each side with $\mm{A}_{lk}$ to finally get
\begin{equation}
    \lambda (\mm{A}^{T} \mm{A}) \bm{b} = (\mm{A}^{T} \mm{B} \mm{A}) \bm{b},
\end{equation}
and thus the product of eigenvalues in this problem is simply
\begin{equation}
    |\mm{H}| = \frac{|\mm{A}^{T} \mm{B} \mm{A}|}{|\mm{A}^{T} \mm{A}|}.
\end{equation}
We see that there are two terms in this expression.
The denominator $|\mm{A}^{T} \mm{A}|$ measures the angles between the gradients of the constraining functions, while the numerator $|\mm{A}^{T} \mm{B} \mm{A}|$ measures the shape and strength of the confining field in these directions.

\section{Calculation for the trimer}

\begin{figure*}[htbp]
    \centering
    \includegraphics[width=0.9\linewidth]{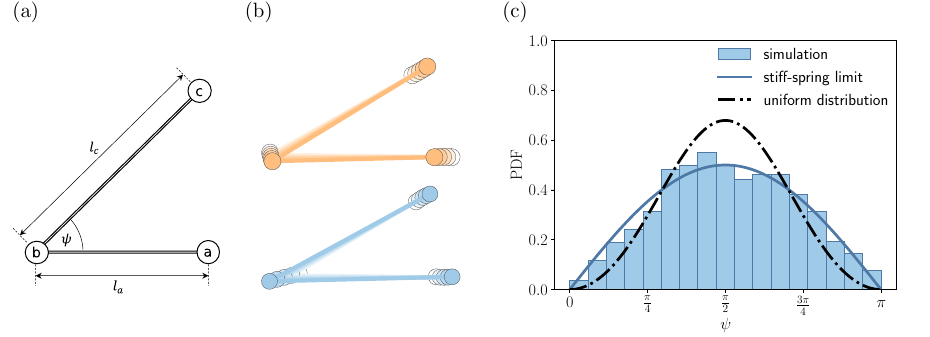}
    \caption{(a) Parametrisation of the trimer uses the position of the central particle ($b$), bond angle $\psi$ and bond lengths $l_a$ and $l_b$, with identical equilibrium spring lengths $l_0$. Rigid-body rotation around the particle $b$ is omitted for clarity. (b) Motion along the symmetric (top) and antisymmetric eigenvectors of the hessian $\mm{H}$ of the harmonic confinement function. Note that the central bead must move for the overall motion to be orthogonal to the constraining manifold $\mathcal{M}$. (c) Equilibrium probability density function (PDF) for the bond angle $\psi$ from the Langevin numerical simulation. Solid line is the theoretical prediction of Eq.~\eqref{eq:trimer_final}\changed{, dashed line is the uniform distibution given by projection of canonical measure given by Eq.~\eqref{eq:uniform_trimer}.}
    }
    \label{fig:trimer}
\end{figure*}

In the aforementioned case of a trimer, depicted in Fig.~\ref{fig:trimer}(a), we have 3 beads located at $\bm{r}_a, \bm{r}_b$ and $\bm{r}_c$ with bond extensions $P_a, P_c$ given by
\begin{eqnarray}
    P_a & = & |\bm{r}_a - \bm{r}_b| - l_0, \\
    P_c & = & |\bm{r}_c - \bm{r}_b| - l_0,
\end{eqnarray}
$l_0$ being the equilibrium bond length, and the confining potential given by  
\begin{equation}
    W(\bm{q}) = P_a^2 + P_c^2.
\end{equation}
We now pick the parametrisation $\bm{q} = \bm{\zeta^*}(\bm{w})$ with $\bm{w} = [x, y, z, \alpha, \beta, \gamma, \psi, l_a, l_c]^{T}$. Cartesian coordinates $(x,y,z)$ describe the  position of the central bead ($b$), $(\alpha,\beta,\gamma)$ are the Euler angles that encode the direction of bead $a$ seen from $b$, and $\psi$ is the bond angle between the edges $ab$ and $bc$. The positions of beads are thus given by
\begin{eqnarray}
    \bm{r}_a & = & [x,y,z]^{T} + l_a \mm{E}_{\alpha\beta\gamma} [1,0,0]^{T},\\
    \bm{r}_b & = & [x,y,z]^{T}, \\
    \bm{r}_c & = & [x,y,z]^{T} + l_c \mm{E}_{\alpha\beta\gamma} [\cos\psi,\sin\psi,0]^{T}.
\end{eqnarray}
We defined the rotation matrix $\mm{E}_{\alpha\beta\gamma}$ in terms of the Euler angles in the Appendix. In addition, we note that $P_a = l_a - l_0$ and $P_c = l_c - l_0$. We compute the metric term as
\begin{equation}
    |\mm{J}^{T} \mm{J}| = \frac{1}{2} l_0^8 \sin^2 \beta \sin^2 \psi \left( 7 - \cos 2 \psi \right),
    \label{eq:uniform_trimer}
\end{equation}
from the matrix $\mm{J}$ listed in the Appendix. 
\changed{Next, we compute the shape term using Eq.~\eqref{eq:matrix_a_defn}. Intermediate steps of the calculation require computing a 2-by-9 matrix $\mm{A}$ which has a number of zero entries but contains lengthy trigonometric expressions. These simplify somewhat for points on $\mathcal{M}$, but were handled with symbolic computation engine Mathematica. Computing 2-by-2 matrices $\mm{A}^{T} \mm{A}$ and $\mm{A}^T \mm{B} \mm{A}$, after simplification, gives compact closed-form expressions, which, as expected, depend only on the angle $\psi$, via}  
\begin{eqnarray}
    \mm{A}^{T} \mm{A}        & = &
                                   \begin{bmatrix}
                                       2         & \cos \psi \\
                                       \cos \psi & 2
                                   \end{bmatrix},
    \\
    \mm{A}^{T} \mm{B} \mm{A} & = &
                                   \begin{bmatrix}
                                       9 + \cos 2 \psi & 8 \cos \psi     \\
                                       8 \cos \psi     & 9 + \cos 2 \psi
                                   \end{bmatrix},
\end{eqnarray}
thus finding
\begin{equation}
    |\mm{H}| = \frac{|\mm{A}^{T} \mm{B} \mm{A}|}{|\mm{A}^{T} \mm{A}|} =  2 (7 - \cos 2 \psi).
\end{equation}
We therefore have
\begin{equation}
    \frac{|\mm{J}^{T} \mm{J}|^{1/2}}{|\mm{H}|^{1/2}} = \frac{l_0^4}{2} \sin\beta \sin\psi, 
\end{equation}
and for the final distribution, we find
\begin{equation}
        \dd p_\infty =  N \sin\beta \sin\psi \,\, \dd x\, \dd y\, \dd z\, \dd \alpha\, \dd \beta\, \dd \gamma\, \dd \psi, \label{eq:trimer_final}
\end{equation}
where $N$ is a normalisation constant. Therefore, the marginal density of the bond angle is proportional to $\sin\psi$ as expected. However, this is simply not the case of a uniform distribution––that distribution corresponds to $|\mm{H}| = \text{const}$ and the notion of \emph{uniformity} is inherited from the canonical measure in the canonical (Cartesian) coordinates \changed{displayed as uniform distribution in Fig.~\ref{fig:trimer}(c)}.
This can be examined in greater detail by looking at the eigenvectors of $\mm{H}$, visualised in Fig.~\ref{fig:trimer}(b). There are two eigenvectors which correspond to symmetric and antisymmetric motion of the terminal beads. Counterintuitively, the vectors in $\mathcal{M}^{\perp}$ change the orientation of at least one bond when the bond length changes.

The simplification lies in the fortuitous cancellation of the $(7-\cos 2\psi)$ term, rather than its absence.
This is best highlighted by an example where no such cancellation occurs, which we present in the next Section.

We additionally corroborate this result with a BD simulation in Python using the package \texttt{pychastic}~\cite{Waszkiewicz_2023}. Choosing $k_{B}T$ as the energy scale, $l_0$ as the distance scale and $l_0^2 / D$, as the timescale,  with $D$ being the diffusion coefficient, the corresponding stochastic BD equation takes the form
\begin{equation}
    \dd\bm{r}_i = - \bm{\nabla} (U + k^2 W) \dd t + \sqrt{2}\,\dd\bm{\mathcal{B}}_i,
\end{equation}
with $\bm{\mathcal{B}}_i$ denoting the standard Wiener process. We have performed $n_\text{traj} = 4000$ simulations with $k = 35$ (in units of $k_B T/l_0$) for a time up to $t_{max} = 10$ with time step $\delta t = 10^{-5}$. The simulation results in Fig. \ref{fig:trimer}(c) complement the prediction of Eq.~\eqref{eq:trimer_final}. We use the same code for a less studied example of a tetramer in the following.

\section{Cyclic tetramer}

\begin{figure*}[htbp]
    \centering
    \includegraphics[width=0.9\linewidth]{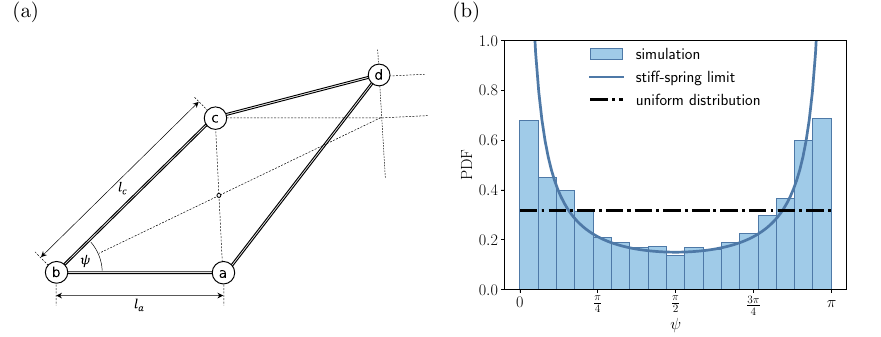}
    \caption{(a) The shape of a cyclic tetramer is specified with the bead $b$ taken as reference point. The directions of sections $ab$ and $bc$ span a rhombus, which is used to define the position of bead $d$. 
    (b) Equilibrium probability density function (PDF) for the bond angle $\psi$ from the numerical simulation.  The solid line is the theoretical prediction of Eq.~\eqref{eq:tetramer_final}. \changed{The dashed line is the uniform distribution obtained by projecting the canonical (Cartesian) measure.}
    }
    \label{fig:tetramer}
\end{figure*}

A more complex, and yet fully tractable analytically, is the case of a cyclic tetramer, a molecule of 4 beads joined into a quadrilateral. For the sake of brevity, we discuss the dynamics in two dimensions. The calculation in 3D is completely analogous but involves much longer intermediate expressions due to the additional rotational degrees of freedom and a dihedral angle.
We pick the parametrisation where the 4 beads are located at $\bm{r}_a,\bm{r}_b,\bm{r}_c$ and $\bm{r}_d$, with their positions given by
\begin{eqnarray}
    \bm{r}_a        & = & [x, y]^{T} + l_a \mm{E}_\alpha [1, 0]^{T},                                                                      \\
    \bm{r}_b        & = & [x, y]^{T},                                                                                                     \\
    \bm{r}_c        & = & [x, y]^{T} + l_c \mm{E}_\alpha [\cos\psi, \sin\psi]^{T},                                                        \\
    \bm{r}_d        & = & [x, y]^{T} + \mm{E}_\alpha \left( \left( 1+ \delta \right) \bm{v}_\delta + \epsilon \, \bm{v}_\epsilon \right), \\
\end{eqnarray}
where $\mm{E}_\alpha$ is a rotation matrix defined in the Appendix, and the vectors $\bm{v}_{\delta,\epsilon}$, written as \begin{eqnarray}
    \bm{v}_\delta   & = & \frac{[l_a + l_c \cos \psi , l_c \sin\psi ]^{T}}{2},                                                            \\
    \bm{v}_\epsilon & = & \frac{[l_a - l_c \cos\psi,-\sin\psi]^{T}}{2},
\end{eqnarray}
point along the diagonals of the rhombus spanned by the edges $ab$ and $bc$. Finally, $\epsilon$ and $\delta$ encode the position of the bead located at $\bm{r}_d$. The four constraining surfaces are given by
\begin{eqnarray}
    P_a & = & |\bm{r}_a - \bm{r}_b| - l_0, \\
    P_b & = & |\bm{r}_b - \bm{r}_c| - l_0, \\
    P_c & = & |\bm{r}_c - \bm{r}_d| - l_0, \\
    P_d & = & |\bm{r}_d - \bm{r}_a| - l_0,
\end{eqnarray}
and $\mathcal{M}$ is given by $l_a = l_c = l_0, \delta = 1, \epsilon = 0$.

The metric term is simply constant with $|\mm{J}^{T}\mm{J}| = 16 l_0^4$, while the shape term is non-trivial.
Denoting $\cos\psi = c_\psi$  for brevity, the shape factors are given by
\begin{eqnarray}
    \mm{A}^{T} \mm{A} &=& 
                   \begin{bmatrix}
                        2 & c_{\psi } & 0 & -c_{\psi } \\
                        c_{\psi } & 2 & -c_{\psi } & 0 \\
                        0 & -c_{\psi } & 2 & c_{\psi } \\
                        -c_{\psi } & 0 & c_{\psi } & 2 \\
                   \end{bmatrix},
    \\    
    \mm{A}^{T} \mm{B} \mm{A} &=& 4
            \begin{bmatrix}
 c_ {\psi }^2+2 & 2 c_{\psi } & -c_ {\psi }^2 & -2 c_{\psi } \\
 2 c_{\psi } & c_ {\psi }^2+2 & -2 c_{\psi } & -c_ {\psi }^2 \\
 -c_{\psi }^2 & -2 c_{\psi } & c_ {\psi }^2+2 & 2 c_{\psi } \\
 -2 c_{\psi } & -c_ {\psi }^2 & 2 c_{\psi } & c_ {\psi }^2+2 \\
            \end{bmatrix},
\end{eqnarray}
yielding $|\mm{H}| = 256 \sin^2 \psi$. As a result, we get that
\begin{equation}
    \dd p_{\infty} \propto \frac{1}{\sin \psi} \dd\psi, \label{eq:tetramer_final}
\end{equation}
which is non-normalizable near 0 and $\pi$. Physically, as we approach the stiff spring limit, the tetramer spends more and more time in the folded state, and unless some repulsive potential is added, the molecule in the stiff spring limit looks predominantly like a trimer, alternating between beads $(a,c)$ and $(b,d)$ coinciding in space.

This surprising result is also seen in the numerical simulation. In this case, a strong repulsive potential was added to prevent bead overlaps for distances less than $0.05 l_0$. The final bond angle distribution shown in Figure~\ref{fig:tetramer}(b) coincides well with $0.15 / \sin \psi $.

\section{Conclusions}

In this work, we have provided a procedure by which one can find the equilibrium probability probability distribution function of the configurations of a bead-spring model of a polymer in the overdamped regime in the limit of infinite stiffness of the bonds. In particular, we have demonstrated that the shape of the confining spring potential persists in the limiting distribution. By rephrasing the problem as a mathematically rigorous limit, we have shown that the classical expression given by Fixman~\cite{Fixman_1974}, later reproduced in numerous works and books, for example in Ref.~\cite{Frenkel_2023}, is missing a critical term describing the shape of the confining potential, $|\mm{H}|$. \changed{The  presented reasoning can be applied without changes to potentials which depend in an eventually monotone way on the large parameter $k$, dropping the assumption of linear dependence on $k$. Furthermore, in the case of a very flat and very sharp confining potentials, which do not admit a second-order Taylor expansion with an almost-surely non-vanishing Hessian, one can sometimes reduce the problem to the presented case by a change variables in the soft coordinates such as $s \to s^\gamma$. For other, badly-behaved confining potentials, such as $1/\log(1/s)$, a different methodology is required.}

\changed{Our work} shows that the reasoning presented by Frenkel \& Smit~\cite{Frenkel_2023} cannot be universally applied. First, the dynamics and the distribution of a polymer connected with (stiff or soft) springs is independent of the mass of the monomers in the overdamped regime, and thus the limiting distribution cannot depend on the mass of the monomers.
Second, putting aside the question of existence of the rigid rod distribution (or the lack thereof), the supposition that the stiff spring limit should always be the same, regardless of the details of the confining potential is clearly not true, and harmonic potentials will lead to a different outcome than springs that realise uniform confinement around the constraining manifold, as previously remarked, e.g., by van Kampen \& Lodder \cite{van_Kampen_1984}, although without a general mathematical formulation. \changed{Thus it cannot be true that the ratio of densities of the rigid rod distribution to the stiff spring limit distribution is given by the manifold $\mathcal{M}$ alone.} The difference \changed{between uniform confinement and harmonic spring confinement} can be arbitrarily large, as shown in the tetramer example.  Frenkel \& Smit remark in their book that for the bond length constraints of the type most often used in MD simulations, the effect of hard constraints on the distribution functions seems to be relatively small~\cite[Chapter 15.1.1]{Frenkel_2023}. We believe this statement to be misleading. In the discussed case of the trimer molecule, the effect might indeed be small, but we have shown that for the tetramer system, which is still a relatively simple setup, the effect is pronounced. To provide practical insight, we propose a technique for the analysis of limiting distributions once the confining potential is specified in detail.

\begin{acknowledgments}
    The Authors thank Bartłomiej Lewandowski for his insightful comments regarding convergence of distributions. 
    The work of ML and RW was supported by the National Science Centre of Poland (FundRef DOI: http://dx.doi.org/10.13039/501100004281) grant Sonata to ML no.
    2018/31/D/ST3/02408.
\end{acknowledgments}

\appendix

\section{Auxiliary matrices used in calculations}

Using the notation $s_\alpha = \sin \alpha$, $c_\alpha = \cos \alpha$ we can compactly express the matrices used.
The 2D rotation matrix is given by
\begin{equation}
    \mm{E}_\alpha =
    \begin{bmatrix}
        c_\alpha & -s_\alpha \\
        s_\alpha & c_\alpha
    \end{bmatrix}.
\end{equation}
The 3D rotation matrix is given by
\begin{equation}
    \mm{E}_{\alpha\beta\gamma} =
    \begin{bmatrix}
        c_{\alpha } c_{\beta } c_{\gamma }-s_{\alpha } s_{\gamma } & \ -c_{\alpha } c_{\beta } s_{\gamma }-c_{\gamma } s_{\alpha } & \ c_{\alpha } s_{\beta } \\
        c_{\beta } c_{\gamma } s_{\alpha }+c_{\alpha } s_{\gamma } & \ c_{\alpha } c_{\gamma }-c_{\beta } s_{\alpha } s_{\gamma }  & \ s_{\alpha } s_{\beta } \\
        -c_{\gamma } s_{\beta }                                    & s_{\beta } s_{\gamma }                                        & c_{\beta }               \\
    \end{bmatrix}.
\end{equation}

For our paremetrisation of the trimer, the Jacobian $\mm{J}$ can be expressed as a block matrix with $3\times4$ blocks $\mm{Q}_i$
\begin{equation}
    \mm{J} =
    \begin{bmatrix}
        \mm{1} & \mm{Q}_1 \\
        \mm{1} & \mm{Q}_2 \\
        \mm{1} & \mm{Q}_3 \\
    \end{bmatrix}.
\end{equation}
Writing $\gamma + \psi = \theta$ for brevity, the blocks $\mm{Q}_i$ are given by
\begin{equation}
    \mm{Q}_1^{T} = l_0
    \begin{bmatrix}
        - s_\alpha c_\beta c_\gamma-c_\alpha s_\gamma & c_\alpha c_\beta c_\gamma-s_\alpha s_\gamma & 0                 \\
        -c_\alpha s_\beta c_\gamma                    & -s_\alpha s_\beta c_\gamma                  & -c_\beta c_\gamma \\
        -s_\alpha c_\gamma-c_\alpha c_\beta s_\gamma  & c_\alpha c_\gamma-s_\alpha c_\beta s_\gamma & s_\beta s_\gamma  \\
        0                                             & 0                                           & 0
    \end{bmatrix}
\end{equation}

\begin{equation}
    \mm{Q}_2^{T} =
    \begin{bmatrix}
        0 & 0 & 0 \\
        0 & 0 & 0 \\
        0 & 0 & 0 \\
        0 & 0 & 0 \\
    \end{bmatrix}
\end{equation}
\phantom{.
}
\begin{equation}
    \mm{Q}_3^{T} = l_0
    \begin{bmatrix}
        -s_\alpha c_\beta c_\theta-c_\alpha s_\theta & c_\alpha c_\beta c_\theta-s_\alpha s_\theta & 0                 \\
        -c_\alpha s_\beta c_\theta                   & -s_\alpha s_\beta c_\theta                  & -c_\beta c_\theta \\
        -s_\alpha c_\theta-c_\alpha c_\beta s_\theta & c_\alpha c_\theta-s_\alpha c_\beta s_\theta & s_\beta s_\theta  \\
        -s_\alpha c_\theta-c_\alpha c_\beta s_\theta & c_\alpha c_\theta-s_\alpha c_\beta s_\theta & s_\beta s_\theta  \\
    \end{bmatrix}
\end{equation}

\bibliography{sources}

\end{document}